\documentstyle[pre,preprint,aps]{revtex}
\begin{document}
\title{Photon transport in thin disordered slabs}
\author{Venkatesh Gopal$^{(1)}$,
S.Anantha Ramakrishna$^{(2)}$,
A.K.Sood$^{(1)}$\footnote{e-mail:asood@physics.iisc.ernet.in}
and N.Kumar$^{(2)}$\footnote{e-mail:nkumar@rri.res.in}}
\address{$^{(1)}$ Department of Physics, Indian Institute of Science, Bangalore 560 012, 
India}
\address{$^{(2)}$ Raman Research Institute, C.V. Raman Avenue, Bangalore 560 080, India}
\maketitle
\widetext
\begin{abstract}

We examine using Monte Carlo simulations, photon transport
in optically `thin' slabs whose thickness L is only a few times
the transport mean free path $l^{*}$, with particles of different 
scattering anisotropies. The confined geometry causes an auto-selection of 
only photons with looping paths to remain within the slab.
The results of the Monte Carlo 
simulations are borne out by our analytical treatment which incorporates 
the directional persistence by the use of  
the Ornstein-Uhlenbeck process, which interpolates between the short time
ballistic and long time diffusive regimes.  


\end{abstract} 
\noindent
\section{Introduction}

The multiple scattering of light as it travels through a random medium
is a problem that is interesting both because of the fundamental
physics that it involves as well as the potential that it offers to
craft novel imaging techniques for use in optically turbid media
\cite{Yodh}. In the limit of a very large number of scattering
events, the transport of the light intensity through the medium can be
described by a diffusion equation which can be shown to arise out of
the equation of radiative transfer \cite{Ishimaru}. In the opposite
limit of only a few scattering events, single scattering theory with
multiple scattering corrections is adequate \cite{Ishimaru}. It is in
the intermediate regime when the transport is neither entirely
diffusive nor largely ballistic that both these pictures prove
insufficient. In a large scattering medium, photons scatter many
times and after a few scattering events they are directionally
randomized resulting in the photon performing a random walk. The
transport mean free path $l^*$ is the length scale over which
directional memory is lost. Typically, the assumption of diffusive
intensity transport is a good approximation when the thickness of the
medium $L$, is greater than about eight times $l^*$ \cite{Pine,Alfano}, the
step size of the photon random walk. The reason for the breakdown of 
the diffusion theory at shorter length scales can be traced to the directional 
persistence of the random photon walks. However, it is of 
importance to study alternatives to the zero-persistence diffusion
theory for use in scattering media whose thickness is only a
few times $l^*$ \cite{Telegrapher,Anantha,Boguna}.

Recently, in experiments that studied the propagation of an ultrashort
pulse of light through an optically `thin', strongly scattering
slab \cite{Kop}, the photon diffusion
coefficient $D$ measured from the temporal behaviour of the transmitted
light was found to decrease as a function of the slab thickness. 
This result is intriguing because the diffusion coefficient should be 
independent of geometry. 
We suspect that it is the persistence of the
random walks of the photons that could cause this apparent reduction of 
the diffusion coefficient and we explore this possibility. 
Monte Carlo simulations of the random walks  naturally account for 
the persistence arising due to constant speed \cite{Anantha}
and finite mean free 
paths of the photons. Analytically, we have attempted to incorporate this
persistence by the use of the Ornstein-Uhlenbeck (OU) process of Brownian 
motion.  

We should note here that we are dealing with the case of light (which is a 
wave) propagating in a random medium, where the randomness may be in space 
(quenched disorder), but treat it as the Brownian motion of a classical particle
i.e., temporal disorder or a Stochastic process. This approximation 
of an incoherent energy transport is 
valid in the weak scattering limit ($kl^{*} \gg 1$, where $k=2\pi / \lambda$). 

In this paper, we describe the results of our Monte Carlo simulations
to study photon transport in thin slabs. We show that in small slabs,
photons that traverse large paths are forced by the constrained
geometry to travel in paths that loop back upon themselves, thus
lowering the rate at which the photons are transported in the medium.
In our effort to find
alternatives to the widely used zero-persistence 
diffusion picture of transport, we
examined the Ornstein-Uhlenbeck theory of Brownian motion. By
making an ansatz, we found the OU process to yield  close
agreement with the results of our simulations. We find the OU process
to adequately describe the ballistic motion at short times and become
identical with the diffusion approximation asymptotically at long
times. 

The paper is organized as follows. In section II, the Monte Carlo
simulation technique used is described. Our results, qualitatively
explaining the reduced photon diffusion coefficient in confined
geometries are presented in section III. In Section IV we describe the
Ornstein-Uhlenbeck process and the modifications made to use it to
describe the transport of multiply scattered light. We also compare
results obtained using the OU process with those from random-walk
simulations of photon transport. Finally, we conclude in section V.

\section{Monte Carlo simulations}
The procedure for our
Monte Carlo simulations was as follows. Photons were launched from
the centre of a slab in the + z direction. The slab was of infinite
transverse extent but with a finite thickness L in the z direction.
We chose the centre of the box as the source of the photons as the
results are then easier to interpret. The simulation modelled photon
transport assuming that the photons travel exponentially distributed
lengths between scattering events. The probability $P(s)$ of
travelling a ballistic path length $s$ is given by the familiar
Lambert-Beer law, $P(s) = exp(-s/l_s)$, where $l_s$ is the scattering
mean free path of the photons in the medium. The scattering mean free
path is set by the scattering cross-section $\sigma$ and the number
density $\phi$ of the scatterers as $l_{s} = 1/\sigma \phi$. The
scattering cross-sections were calculated using Mie theory
\cite{Bohren}. The random paths between scattering events were
generated taking $s = l_s \cdot \ln(\mathcal R)$, where $\mathcal R$
is a random number uniformly distributed between 0 and 1
\cite{Numrec}. The scattering angles were chosen such that they had a
distribution of directions given by the Henyey-Greenstein phase
function, where the probability of scattering at an angle $\theta$
relative to the incident direction of the photon is given by
\begin{equation} P(\cos \theta) = \frac{1 - g^2}{(1 + g^2 - 2g \cos
\theta)^{3/2}} \end{equation} and where $g = \langle \cos \theta
\rangle = 1 - (l_s/l^*)$ is the scattering anisotropy. Thus, after
being launched from the centre of the slab, photons were repeatedly
scattered until they encountered a boundary, at which point they were
assumed to have escaped from the slab and a new photon was launched
from the centre. Each such random walk trajectory generated was
stored. Time dependent statistics were calculated by dividing these
trajectories into numerous slices of equal lengths and hence equal
intervals of time. Simulations were performed for three different
scattering anisotropies (g = 0.06, 0.423 and 0.732) to study the range
from nearly isotropic to highly anisotropic scatterers. 

As a check on the accuracy of our simulations, the diffuse
transmission probability \cite{Durian1} as a function of the slab
thickness for slabs thicker than 5$l^*$ and the photon diffusion
coefficient $D_{0} = cl^{*}/3$ were calculated. Excellent agreement
with theoretical predictions was obtained in both cases.

\section{Photon diffusion in confined geometries}
 We begin by providing a
qualitative description of the of the motion of the photon cloud at
short times, which we then support with data from our simulations.
Consider a source of photons at the centre of a slab of scattering
material and being launched in the +z direction. A characteristic
time scale in the problem is the time $t_{b}= L/c$, that it would take
an unscattered or ballistic photon to traverse the thickness of the
slab. At very short times, $(t < t_{b})$, there are two kinds of
photons, those that are ballistic and have not yet been scattered and
those that have. There is a large separation in the rate at which
these photons are moving away from the source. At long enough times,
gradually, all photons suffer scattering events and a smooth
distribution of distances from the source is obtained, but at short
times there is a sharp distinction between ballistic and multiply
scattered photons. 

If now the scattering takes place in a confined geometry, before a
smooth distribution of radial distances from the origin is allowed to
form, then a photon that must multiply scatter and {\it yet} stay
within the confines of the scattering medium must necessarily follow a
tortuous path that loops around and `doubles back' on itself. As we
show later, this leads to a reduction of our diffusion coefficient,
for if the photon does not `loop back', it exits the slab along with
or just after the ballistic pulse. For these paths that loop back
upon themselves, the rate at which the radial distance from the
origin, a `front velocity' of the photon cloud as it were, is clearly
increasing at a much slower rate than it would in an unconfined
medium. On increasing the size of the slab though, the number of
ballistic photons decreases exponentially and the well known smooth
path length distribution from the origin for diffusion is obtained
\cite{Pine}. Figures 1 and 2, which we now describe, support this
picture. 

In the random walk simulations, we probed the rate of transport of
photons in the medium by investigating the mean square displacement of
the photons as a function of time. This is reflected in the rate at
which the photon cloud in the medium expands, the mean square
displacement simply being the mean radius of the photon cloud.
Photons were propagated in slabs of varying thicknesses as described
previously. Some of the results are shown in Figure 1. In both
figures we show the local slope of the mean square displacement
denoted by an effective diffusion constant $D(t;L)$ normalized to
$6D_{0}$, as a function of time. At long times, and for large slabs,
we have the well known Einstein result, 
$\langle \Delta r^{2} \rangle= 6D_{0}t$,
where $D_{0} = cl^{*}/3$ is the photon diffusion coefficient in an
infinite medium, and the local slope thus tends to unity. Fig. 1(a)
shows the variation of this local slope for nearly isotropic
scatterers with a scattering anisotropy $g = 0.06$ for different slab
thicknesses, while Fig. 1(b) shows the same data for anisotropic
scatterers with $g = 0.732$. All the curves show a sharp break marked
by a symbol. This is the point at which the ballistic photons escape
from the slab. This results in a sharp decrease in the mean square
displacement leading to an artificial negative value for the diffusion
coefficient, which is avoided by showing the slopes just before and
after the escape of the ballistic pulse. 
In both figures 1(a) and
1(b), the circles denote slabs of thickness 2$l^{*}$, the triangles
slabs of 5$l^{*}$ and the squares 10$l^{*}$. 

The main features of Fig. (1) are that at short times, photon motion is
ballistic and the mean square displacement increases quadratically in
time ({\it i.e} $D(t;L) \propto t$). At long times, for large slabs,
the motion is diffusive and the local slope is unity. After most
photons have escaped from the slab, the only surviving photons in the
medium are those that are confined to the transverse plane and the local
slope is that corresponding to two dimensional diffusion, {\it i.e.}
$D(t \gg t_{b}) = 2D_{0}/3$. At long times it can be seen that the
normalized local slope tends to 2/3.

To show that the multiply scattered photons in the medium undergo long
circular scattering loops, we investigated the normalized velocity
autocorrelation function (VACF) for photons in the medium, which are
shown in Fig. 2(a) ($g = 0.06$) and Fig. 2(b) ($g = 0.723$). The
calculation of the VACF $C(t) = \langle {\bf v}(0) \cdot {\bf v}(t)
\rangle /\langle {\bf v}(0) \cdot {\bf v}(0) \rangle$ is performed as
follows. Each photon trajectory is divided into a number of sections
corresponding to the resolution in time that is chosen. At the origin
of each such `time step', the angle made by the path at that point with
the global coordinate system, which is equivalent to the angle made by
the velocity vector, is stored. The correlation function $C(t)$ is now
formed and averaged over all random walk trajectories. Once again the
symbols denote the points at which the ballistic pulse exits the slab
and the circles, triangles and squares represent the same slab
thicknesses as in figure 1. As soon as the ballistic pulse leaves the
medium we see that the VACF is negative, indicating that the photons
have {\it on average}, reversed their direction of motion, exactly what
would be expected for photons travelling such 'loop-like' paths as we
have stated. That the average is influenced shows that we are observing
a feature common to a majority of the photons and not just the reversal
of a few trajectories. This is to be contrasted with the VACF observed
for large slabs where the function is {\it always} positive and smoothly
decays to zero.

In Figure 3, we plot the value of the
normalized local slope $D(t;L)$ just after the escape of the ballistic
pulse. We believe that since this quantity is indicative of the rate at
which photon transport occurs in the medium, it must be reflected in a
pulse transmission experiment in the rate at which photons are
transported across the boundary of the slab.
The resulting figure is
qualitatively similiar to  the data presented in \cite{Kop}. It should
however, be noted that we derive the diffusion coefficient $D$ from the 
behaviour occuring over relatively short times ($<10t^{*}$), whereas, in Ref.
\cite{Kop}, the decay of the transmitted pulse occurs over longer times 
( upto $100 t^{*}$). Thus, it appears as if the two phenomenoa are different, occuring over different time regimes. The effect of the short time behaviour 
on the overall fitting of the data in Ref.\cite{Kop}, however, cannot be 
ignored.   

While it is well understood that accounting for reflections at the
sample boundary is of great importance in the application of the
diffusion approximation \cite{Adlag,Kaplan,Ospeck,Zhu}, it must be
pointed out that to keep details to a minimum, we have not accounted for
interfacial reflections in our simulations. The effect of a boundary
reflectivity would be to re-inject a small part of the escaping pulse
back into the medium and thus create a small number of photons that
would have even larger residence times. 

\section{The Ornstein-Uhlenbeck process} 
The Ornstein-Uhlenbeck process describes the stochastic behaviour of the
velocity of a Brownian particle. The motion of the particle is
described by the stochastic Langevin equation $\ddot{\vec{X}}+\beta
\dot{\vec{X}} = \vec{A}(t)$ \cite{Chandrasekhar}. The friction
coefficient $\beta$ and the components of the concomitant Gaussian white
noise $\vec{A}(t)$ are related as $\langle A_{i}(t)A_{j}(t + \tau)
\rangle =\frac{2\beta k_{B} T}{m}\delta (\tau)\delta_{ij}$ so as to be
consistent with the condition of thermal equilibrium with the thermal
bath at temperature $T$, where $k_B$ is the Boltzmann constant and m is
the mass of the particle. The quantity relevant to our purposes is the
probability density $P_{\infty}(\vec{r},t;\vec{r_0},\vec{u_0},t=0)$ such
that $P_{\infty}(\vec{r},t;\vec{r_0},\vec{u_0},t=0) d^{3} r$ is the
probability of finding the particle in the volume element $d^{3}r$ at a
position r at time t; given that the particle was at $\vec{r_0}$ with a
velocity $\vec{u_0}$ at a time $t_0(=0) < t$. The
$P_{\infty}(\vec{r},t;\vec{r_0},\vec{u_0},t=0)$ for an infinite medium
is given by \cite{Chandrasekhar}
\begin{equation}
\label{pinfty}
P(\vec{r},t;\vec{r_0},\vec{u_0},t=0) = \left[ {\frac{m \beta^{2}}
{2 \pi k_B T f(t)}} \right]^{3/2}  \exp{\left[ - \frac{m \beta^{2}}
{2 k_B T f(t)} \left| \vec{r}
-\vec{r_0} - \frac{\vec{u_0}}{\beta}(1 - e^{-\beta t}) \right|^{2} \right]}
\end{equation}
where $f(t) = 2\beta t - 3 + 4 e^{-\beta t} -e^{-2 \beta t} $.

The basic motivation to adapt this process to light transport is the
following. The diffusion approximation (Wiener process) is valid in the
limit when $l^* \rightarrow 0$ and $c \rightarrow \infty$ such that the
diffusion coefficient $D_0 = cl^*/3$ is a constant. Multiply scattered
photons on the other hand have a finite mean free path in the medium and
are scattered only after intervals of ballistic flight. 
Some alternatives including a generalized Telegrapher equation\cite{Telegrapher}
have been proposed to account for the transition from ballistic motion 
to diffusive transport. It should be noted that such a 
generalization to higher dimensions(i.e. merely by replacing
$\partial^{2}/\partial x^{2}$ with $\nabla^{2}$), however,
is not quite correct, and certainly not correct for short times when ballistic 
propagation 
dominates over diffusion. Indeed, it has also been shown that
such a generalized
Telegrapher equation provides no better an approximation than the
diffusion theory at short time and length scales in higher
dimensions by comparing with Monte Carlo simulations \cite{Porra}. 
The OU process accounts for a finite speed of propagation 
by making a more physical assumption for Brownian
particles, that of assuming a speed distribution for the particles with
a well defined mean speed, thus avoiding the unphysical infinite
velocity  built into the diffusion equation and
therefore interpolating between the short time ballistic motion ($
\langle \Delta r^{2} \rangle \sim t^{2}$) and the long time diffusive limit ($
\langle \Delta r^{2} \rangle \sim t$ ). Also the particle retains
directional memory for a time $\sim t^* (=\frac{c}{l^*} )$ and the
effects of a finite $l^*$ are thus accounted for. This makes it
attractive to consider the OU process to obtain a transport equation
that would describe transport with a finite mean free path and finite
speed and hence capable of describing both the ballistic and diffuse
components.

Hence we force the following identifications as an ansatz and equate
the root mean square (r.m.s.)  speed and the initial speed $u_0$ of the
photons, to c, the speed of light in the medium. 
We recognize, of course,  that there is no thermal bath nor inertia for light, 
and that equating the 
r.m.s.  speed of the photons to the speed of light in the medium is 
a formal identification forced on us by the O-U stochastic process 
introduced here to approximate the diffusion of light.
Thus our ansatz
becomes $\langle \vec{u}^2 \rangle = c^2 = \frac{3 k_B T}{m}, D =
\frac{cl^*}{3} = \frac{ k_B T}{m \beta }$, and $\frac{\left| \vec{u_0}
\right|}{\beta} = \frac{c}{\beta} = l^*$ Rewriting the probability
distribution (\ref{pinfty}) using these we get
\begin{equation}
P_{\infty}(\vec{r},t;\vec{r_0},\vec{u_0},t=0) = \left[ {\frac{3 }
{2 \pi {l^*}^{2} f(t)}} \right]^{3/2}  \exp{\left[ - \frac{3 }
{2 {l^*}^{2} f(t)
} \left| \vec{r}
-\vec{r_0} - {l^*}\hat{n}(1 - e^{-ct/l^{*}}) \right|^{2} \right]}
\end{equation}
It should be noted that in modelling the random walk of the photons by
the OU process, we have implicitly allowed the speed of the photon to
fluctuate as it propagates through the medium. It should also be
remarked here that some of the attempts made to enforce the strong
constraint of constant photon speed have had only partial success, in
that they could calculate the probability distribution subject to the
speed constraint only in the weaker (average) sense {\it i.e.,}
$\int_{0}^{t} \left[ {\left( \frac{d\vec{r}}{dt} \right)}^{2} -c^{2}
\right] dt = 0 $ \cite{Perelman}. A path integral approach shows
that the finite r.m.s speed defined by the fluctuation-dissipation
theorem for the OU process is a stronger global constraint than 
the speed constraint imposed in Ref.\cite{Perelman}.
 An exact theory for the photon diffusion-at-a-constant-speed
 with a locally fixed speed was recently formulated by us \cite{Anantha}, 
where the deficiencies of the zero-persistence diffusion theory were overcome.
But we could obtain only approximate analytical solutions and numerical exact 
solutions.  In this context, the simplicity of the analytical 
solutions to the OU process
which also incorporates some persistence in the velocity 
space makes it important.

The probability distribution function for a finite slab with absorbing
boundary conditions at z = 0 and z = L is approximately expressed in
terms of $P_\infty$ by the method of images \cite{Morse,Feller} yielding
\begin{equation}
\label{image}
P_{L}(\vec{r},t;\vec{r_0},\vec{u_0},t=0) =  \sum_{n=-
\infty}^{+\infty} \left[ P_{\infty}(\vec{r}+2 n L \hat{z},t;z_0
\hat{z},c \hat{z},t=0) -P_{\infty}(\vec{r}+2 n L \hat{z},t;-z_0 \hat{z},
-c \hat{z},t=0) \right]
\end{equation}
This solution for absorbing boundaries holds only approximately in the
limits which we explain below. The equation for the marginal
probability distribution for the position does not remain invariant when
the initial velocity $\vec{u_0}$ and the initial position $\vec{r_0}$
are changed. More physically, for diffusive motion, {\it all} paths
have an equal probability of occurrence and it is for this reason that
the method of images can be applied. In the OU process on the other
hand, paths retain for a time $t^*$, due to inertia, a `memory' of their
initial direction. Thus at times short compared to the randomization
time $t^*$ or when the distance between the source and the absorbing
boundary is less than the transport mean free path, the method of images
is not strictly valid as there is an imperfect cancellation of forbidden
photon paths and their mirror images. This error however decreases
exponentially with increasing slab thickness and with increasing time.
Also, as can be seen in our results, the errors are small enough to be
neglected for the thicknesses we have considered ($L \geq 2l^{*}~$) with
the source located at the centre of the slab. A rigorous solution but
for a semi-infinite half space only with a single absorbing boundary is
given in Ref.\cite{burkhardt}. The series (\ref{image}) is absolutely
and uniformly convergent. The normalization $\int
P_{L}(\vec{r},t;\vec{r_0},\vec{u_0},t=0) d^{3}r $ reduces with time,
corresponding to the flux of probability density which leaks out of the
slab.

In the diffusion approximation, one imposes the absorbing boundary
conditions not at the physical boundaries but at extrapolated boundaries
at a distance $z_e = 2l^{*}/3$ outside the slab \cite{Lemieux,Morse}.
At times long compared to the `randomization time' ($t > t ^{*}$), the
solution should match with the diffusion approximation. However, for
short times ($t \sim t^{*})$, the photons are ballistic and traverse
only the true thickness of the medium. In the absence of a
comprehensive theory for the boundary position, we adopt the following
interpolation scheme. The extrapolated boundary is kept at the physical
boundary at short times upto $t= t^{*}$ after which it is smoothly moved
to $z_e$ outside the physical boundary asymptotically as $t \rightarrow
\infty$, giving an effective slab thickness of
\begin{equation}
L_{eff} = L + 2 \theta (t-t^{*})\left[ 1 - \exp\{-(t/t^{*}-1)\} \right] z_e
\end{equation}
where $\theta$ is the Heaviside step function. It should be noted that
a fitting parameter of the order of unity could have been used to
determine the time at which the boundary starts to move. However, only
a qualitative understanding is being attempted and such a parameter is
unnecessary.

Figure 4 encapsulates the the effectiveness of the use of the OU
process. Fig. 4(a) shows graphically the effect of the various
boundary conditions. The circles represent the number of photons in the
slab as a function of time, normalized to 1, obtained from Monte Carlo
simulations for a slab thickness of $L = 2l^{*}$ for nearly isotropic
scatterers with $g = 0.06$. The curve marked `a' shows the result when
the extrapolated boundaries are maintained at the physical boundaries of
the cell. As can be seen, while this curve approximately captures the
time at which photons begin to escape from the cell and the photon
number density begins to reduce, it completely fails to fit the long
time diffuse tail. The curve marked `b', is one in which the
extrapolated boundaries are held at the extrapolation length $z_{e} =
2l^{*}/3$ throughout. Here, there is excellent agreement at long times
with the Monte Carlo data but the pulse exits the medium much later than
the ballistic pulse, a consequence of the ballistic pulse having to
traverse a medium whose thickness is $L + 2z_{e}$. The solid line is
the result of our moving boundary conditions which fits curve `a' almost
exactly at short times and agrees very well with the Monte Carlo data at
long times. It is to be noted though that even when the extrapolation
length is set to zero, the ballistic pulse exits the slab faster than a
true ballistic pulse would do. This is a consequence of the fact that
we model the photons as having a distribution of speeds. As a result,
there are photons that are travelling with a speed greater than the
speed of light in the medium resulting in this artefact of a
`pre-ballistic' pulse. As the slab thickness is increased this effect
becomes vanishingly small since there are almost no ballistic photons in
the medium. However, it is important to appreciate that the OU process
describes most of the essential features of the simulation at short
times which would not be possible using the diffusion approximation. 

Figure 4(b) compares the results obtained for the mean square
displacement of the photons from the point at which they are launched,
as a function of time, for a cell of thickness $L = 2l^{*}$. At short
times the transport is predominantly ballistic and the mean square
displacement shows the characteristic quadratic behaviour. The kink in
the curves occurs when the ballistic photons exit the slab. At this
point, the fastest moving photons are lost and thus the average value of
the mean square displacement is sharply lowered. The OU process
compares well with the Monte Carlo data.
Figure 4(c) shows the same data but for a cell
whose thickness $L = 8l^{*}$. Now the regime is one where the diffusion
approximation is valid and excellent agreement is obtained between the
OU process and Monte Carlo data. 

Thus, despite the fact that the method of images is not strictly valid
for early times, we find that the OU process proves reasonably effective
in capturing most of the features of photon transport in confined
geometries, a fact that we feel deems it worthy of further study. 

\section{Conclusions} 
We have, using Monte Carlo simulations of random walking photons,
investigated the effect of a confined geometry on the conversion of
ballistic motion to diffusive transport in a multiply scattering medium.
We find that while most photons exit the medium unscattered or scattered
but nearly undeviated from their original direction, a small fraction
of the photons are forced to travel along paths that close upon themselves.
 Due to the directional persistence of the random photon walks
and the confined geometry, there
is an autoselection of only photons which are effecively travelling backwards
to remain within the slab. 
This looping motion is clearly visible in the velocity autocorrelation
function $C(t)$, where one observes a negative dip in the correlation
function indicating a reversal of direction. This reversal of direction
retards the mean rate at which these multiply scattered photons are
transported through the medium. This reduced mean rate of transport could be
reflected in pulse transmission experiments as an apparent reduction of
the photon diffusion coefficient. 
Thus we find a breakdown of the pure diffusion model in thin slabs and
at short times scales ($\sim l^{*}/c$). The reason being that the mean
square displacement is, on these time scales, necessarily quadratic in
time (i.e. ballistic transport), unlike the long time behaviour which
is diffusive. In this work we have examined the effect of the
cross-over from the short time ballistic to the long time diffusive motion on
photon transport by defining an effective diffusion constant based on
the ensemble of trajectories that remain within the medium (sub-ensemble
average) until a time t. 

We have also investigated the suitability of modelling photon transport
by the Ornstein-Uhlenbeck process.  
We have
     attempted to use the {\it OU process} in contrast to a {\it Wiener process}
     with an unbounded speed so as to incorporate the effects of a
     finite mean free path and hence, persistence in the random photon
     walks, while also implementing the constancy of the speed of the photon
     in a weak average sense. We have also proposed approximate solutions
of the OU process for absorbing boundaries based on the mirror image 
method. 
This yields very good
agreement with data obtained from our random-walk simulations.
In view of the simplicity and accuracy
of the approximate soltuions for absorbing boundaries, this should prove to
be an important and 
useful alternative to the diffusion equation in the intermediate
scattering regime when $2l^{*} \leq L \leq 8l^{*}$. 
Finally, we would like to clarify the underlying idea of our approach. 
In all such treatments of the problem of wave propagation in
     random media as a random walk problem (which is valid under
     certain conditions) one chooses a {\it process} which models the
     stochastic, kinematic aspects as closely as possible.
As we have pointed out earlier, we are not implying
     a literal, naive acceleration/deceleration of a ``massive photon'',
     but only that modelling it as a OU process imposes a finite mean
     free path and a finite speed of the photon in the problem.  The
     OU process manages to impose the constraint of a fixed speed of
     the photon in the sense of a weak global constraint.
Further work is in
progress to extend these ideas to Diffusing wave spectroscopy in
optically `thin' samples.

\acknowledgements

We thank the Supercomputer Education and Research Centre (SERC) at the
Indian Institute of Science for computational facilities. AKS thanks
the Raman Research Institute for a visiting professorship. SAR would
like to thank Prof. Rajaram Nityananda for very useful discussions.

\begin{center}
{\bf Figure Captions }
\end{center}

\noindent
{\sf Figure 1} \\
Variation of the normalized local slope of the mean square displacement
$D(t;L) = \frac{1}{6D_{0}}\frac{d<\Delta r^{2}>}{dt}$ with time for
different values of the optical density $L/l^{*}$. The x-axis is in
units of the randomization time $t^{*} = l^{*}/c$. In the thinnest
slab, very few photons survive at long times and thus the statistics are
very poor. This results in large fluctuations in the local slope of the
mean square displacement. To avoid cluttering the rest of the figure we
have fitted a smoothing spline curve to the noisy data. This brings out
the fact that at long times the photons are effectively performing a
random walk in two dimensions and $D(t;L)$ converges to $0.66$ (denoted
by the dash-dotted line). The symbols correspond to the following slab thicknesses: $\circ$ - $2l^{*}$, $\bigtriangledown$ - $5 l^{*}$, $\Box$ - $10 l^{*}$.

\noindent
{\sf Figure 2}\\
Normalized velocity autocorrelation function as a function of normalized
time $t^{*}$, for two values of the scattering anisotropy.\\

\noindent
{\sf Figure 3}\\ 
The value of the normalized local slope $D(t;L)$ just prior to the
escape of the ballistic pulse is taken to approximate the rate of
diffusion in the medium and is plotted as the effective diffusion
constant for slabs of varying thicknesses. The slab thicknesses have
been scaled by the transport mean free path $l^{*}$ \\

\noindent
{\sf Figure 4}\\ 
Comparison of results obtained by modelling photon transport using the
Ornstein-Uhlenbeck process with those obtained from random walk
simulations. Fig. 4(a) compares the rate at which photons exit the
slab when different boundary conditions are applied. The lower figures
compare the mean square displacement calculated by integrating equation
(4) with the random walk simulations for different slab thicknesses.
Fig. 4(b) shows a thin slab where the diffusion approximation is not
valid while Fig. 4(c) is for a slab where the transport is mainly
diffusive. \\

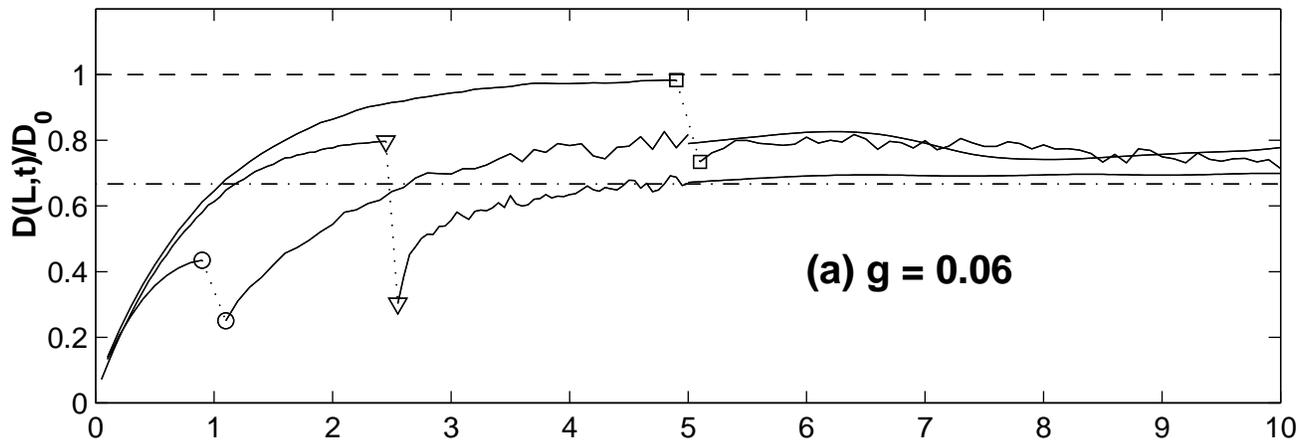
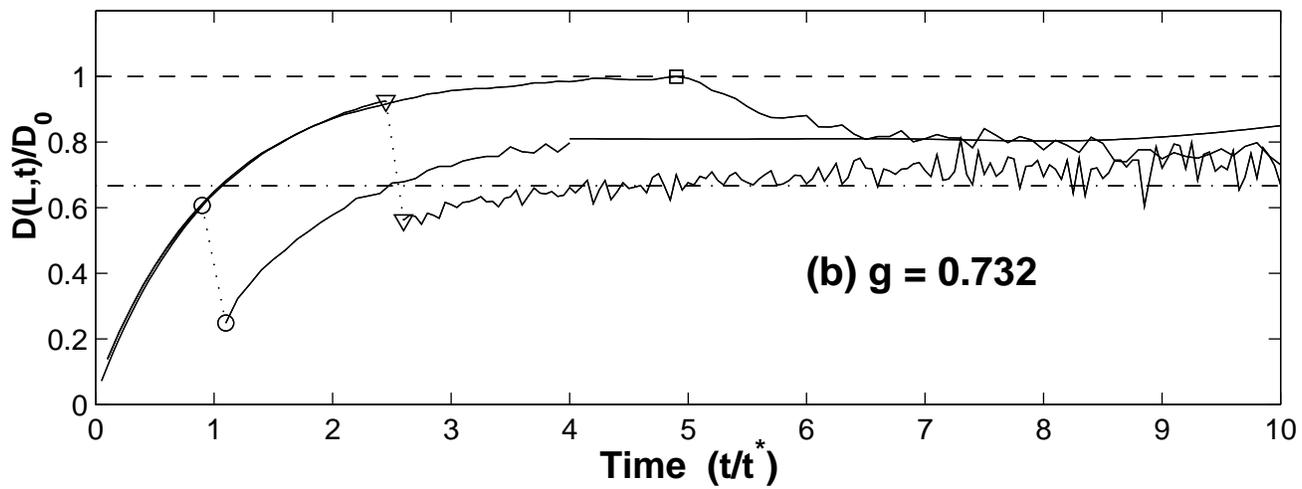
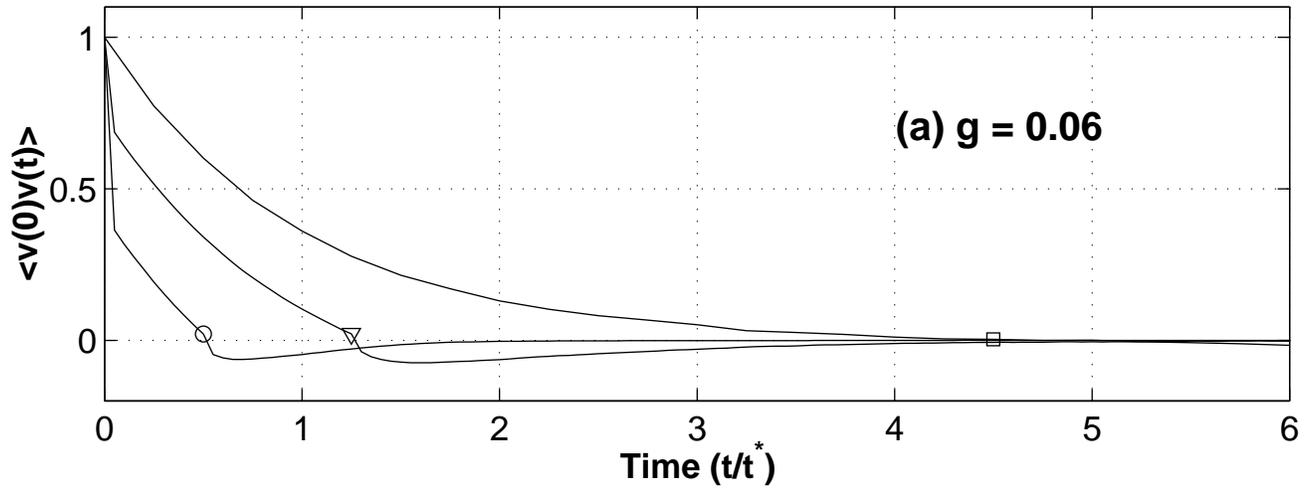
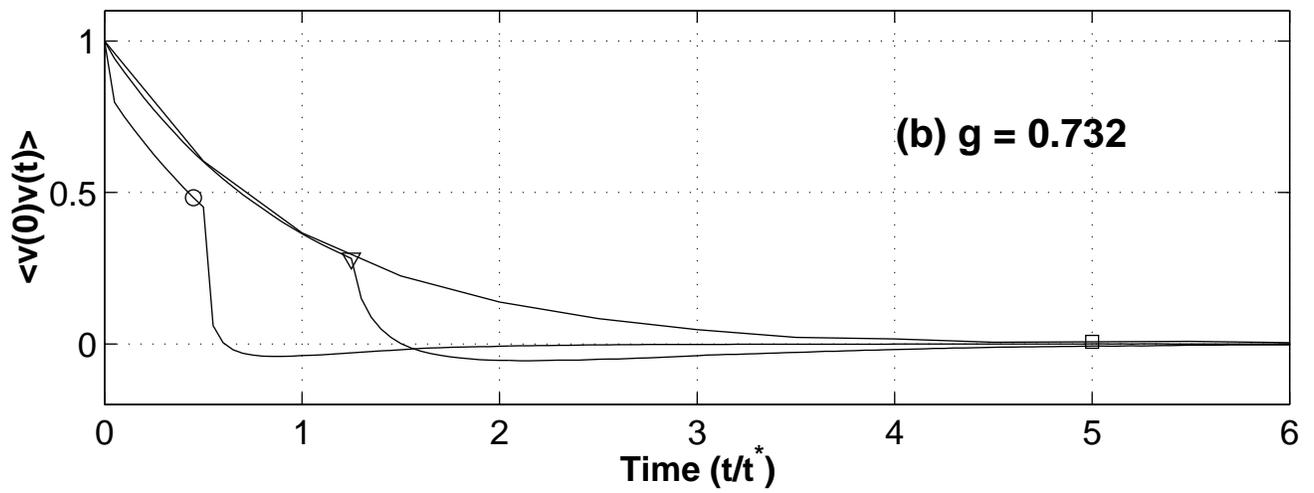
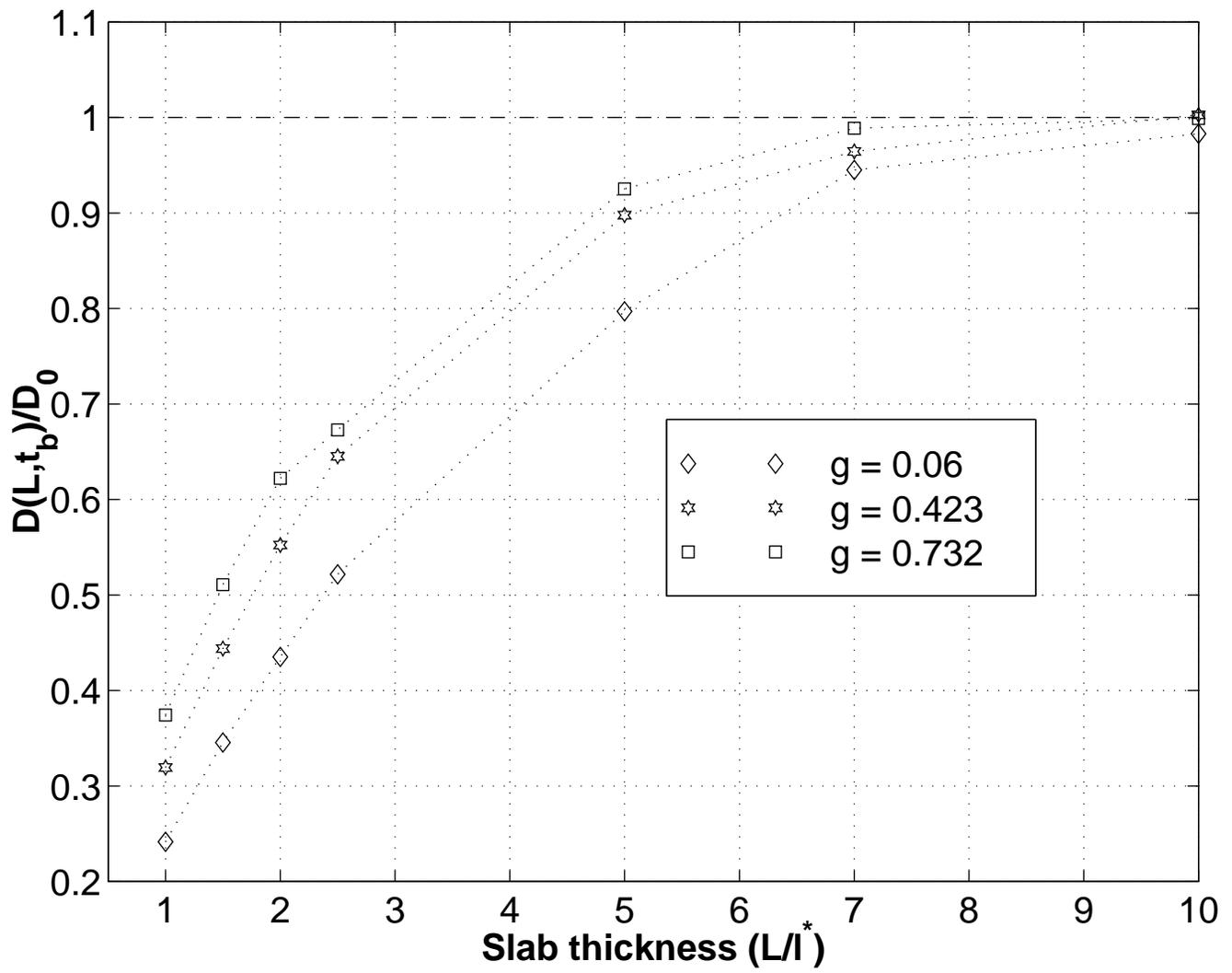
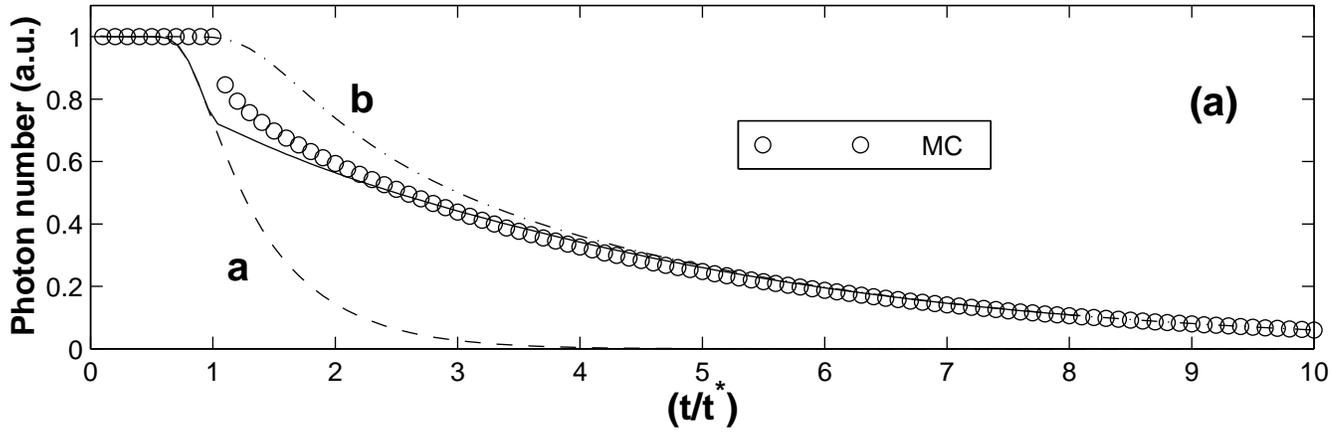
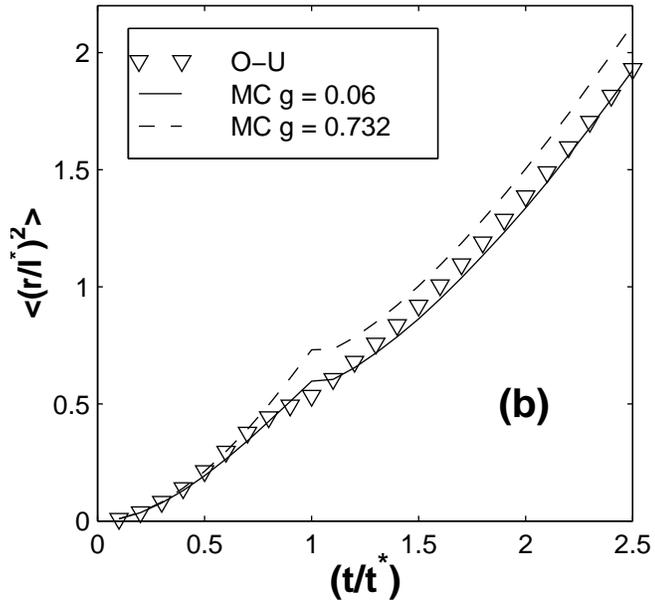
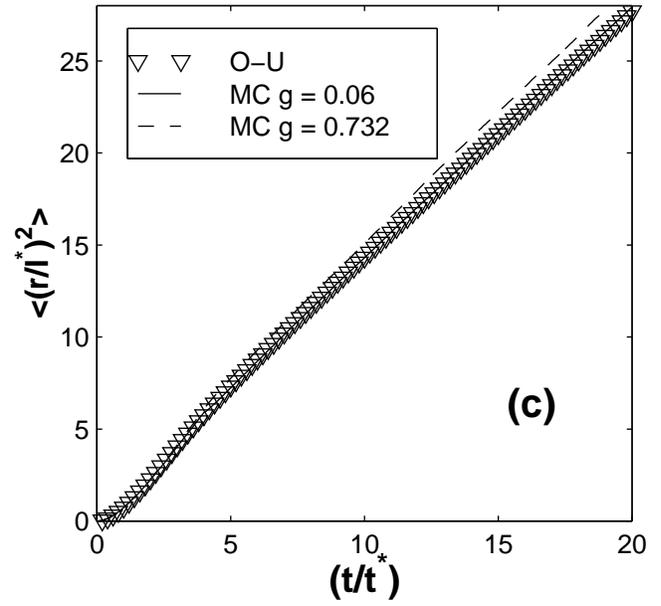
\end{document}